\documentclass[aps,prd,twocolumn,groupedaddress]{revtex4-1}

\usepackage{graphicx}
\usepackage{epstopdf}

\usepackage{textcomp}
\usepackage[utf8]{inputenc}
\usepackage[english]{babel}
\usepackage[breaklinks=true]{hyperref}
\usepackage{color}

\begin{document}

\title{Effect of discrete breathers on macroscopic properties of the Fermi-Pasta-Ulam chain}

\author{Elena~A.~Korznikova$^{1,2}$}
\email{elena.a.korznikova@gmail.com}
\author{Alina~Y.~Morkina$^2$}
\email{alinamorkina@yandex.ru}
\author{Mohit~Singh$^3$}
\email{mohitsingh1997@gmail.com}
\author{Anton~M.~Krivtsov$^{4,5}$}
\email{akrivtsov@bk.ru}
\author{Vitaly~A.~Kuzkin$^{4,5}$}
\email{kuzkinva@gmail.com}
\author{Vakhid~A.~Gani$^{6,7}$}
\email{vagani@mephi.ru}
\author{Yuri~V.~Bebikhov$^{8}$}
\email{bebikhov.yura@mail.ru}
\author{Sergey~V.~Dmitriev$^{1,9}$}
\email{dmitriev.sergey.v@gmail.com}

\affiliation{
$^1$Institute of Molecule and Crystal Physics, Ufa Federal Research Centre of the Russian Academy of Sciences, Ufa 450054, Russia\\
$^2$Ufa State Aviation Technical University, Ufa 450008, Russia\\
$^3$Indian Institute of Technology Kharagpur, Kharagpur 721302, India\\
$^4$Peter the Great Saint Petersburg Polytechnical University, Polytechnicheskaya Street 29, Saint Petersburg, Russia\\
$^5$Institute for Problems in Mechanical Engineering RAS, Bolshoy pr.\ V.O.\ 61, Saint Petersburg, Russia\\
$^6$Department of Mathematics, National Research Nuclear University MEPhI\\ (Moscow Engineering Physics Institute), Moscow 115409, Russia\\
$^7$Institute for Theoretical and Experimental Physics of National Research Centre ``Kurchatov Institute'', Moscow 117218, Russia\\
$^8$North-Eastern Federal University, Polytechnic Institute (branch) in Mirny, 678170 Mirny, Sakha (Yakutia), Russia
\\
$^9$Institute of Mathematics with Computing Centre, UFRC, Russian Academy of Sciences, Ufa 450008, Russia
}%


\begin{abstract}
The effect of discrete breathers (DBs) on macroscopic properties of the Fermi-Pasta-Ulam chain with symmetric and asymmetric potentials is investigated. The total to kinetic energy ratio (related to specific heat), stress (related to thermal expansion), and Young's modulus are monitored during the development of modulational instability of the zone boundary mode. The instability results in the formation of chaotic DBs followed by the transition to thermal equilibrium when DBs disappear due to energy radiation in the form of small-amplitude phonons. It is found that DBs reduce the specific heat for all the considered chain parameters. They increase the thermal expansion when the potential is asymmetric and, as expected, thermal expansion is not observed in the case of symmetric potential. The Young's modulus in the presence of DBs is smaller than in thermal equilibrium for the symmetric potential and for the potential with a small asymmetry, but it is larger than in thermal equilibrium for the potential with greater asymmetry. Our results can be useful for setting experiments on the identification of DBs in crystals by measuring their macroscopic properties.
\end{abstract}

\maketitle


\section{Introduction}
\label{Introduction}

Discovery of discrete breathers (DBs), or intrinsic localized modes (ILMs) --- spatially localized large-amplitude oscillatory modes in nonlinear lattices free of defects --- three decades ago~\cite{1,2,3} has triggered extensive studies devoted to the phenomenon of vibrational energy localization~\cite{19,20}. Experimentally DBs have been excited in the physical systems of different nature, including macroscopic periodic systems~\cite{Mag,pendula,SM}, nonlinear metamaterials, e.g., granular crystals~\cite{Gran0,Gran1,Gran2,Gran4,Gran3,Gran5,Gran7,Gran6} and arrays of micromechanical cantilevers~\cite{Cant1,Cant3,Cant2}, electrical~\cite{ElLatt3,ElLatt2,ElLatt1} and optical lattices~\cite{NonlinOptics}, superconducting Josephson junction arrays \cite{Josephson2,Josephson1}, etc. DBs can be found in crystal lattices~\cite{UFN}, as confirmed by measuring vibrational spectra for alpha-uranium~\cite{UheatCapac,Uranium1,UthermalExp}, helium~\cite{Helium}, NaI~\cite{NaI1,NaI2}, graphite~\cite{Graphite}, and PbSe~\cite{PbSe}. On the other hand, these experimental results in some cases can be interpreted in different ways and they are still debated~\cite{Sievers}.

Nowadays, numerical simulations are very important for investigation of DB properties in various types of crystals. Using {\em ab initio} simulations, the presence of DBs in strained graphene and graphane has been confirmed~\cite{AbInitgraphane,AbInitgraphene}. With the help of molecular dynamics method DBs have been studied in the ionic crystals~\cite{NaIDBMD1,NaIDBMD2,NaIDBMD3}, lattices with pair-wise potentials~\cite{Morse1,Morse2,Morse3}, crystals with covalent bonding~\cite{SiGe,Diamond}, metals~\cite{M1,M2,M2a,M5,M6,M4,M3,M3100,M3101}, intermetallic compounds~\cite{Morse3,OA2,OA0,OA1}, carbon and hydrocarbon nanomaterials~\cite{CH1,CH3,CH2,CH4,CH5,CH6,CH8,CH7,CH9,CH10,CH11,CH13,CH12,CH120,CH121}, $h$-BN~\cite{CH14}, and DNA~\cite{Protein2,Protein1,Protein3,Protein0}.

It is of great importance to understand how DBs affect macroscopic properties of crystals~\cite{Manley}. In the experimental studies, the connection of anomalies in thermal expansion \cite{UthermalExp} and heat capacity \cite{UheatCapac} of $\alpha$-uranium at high temperatures to excitation of DBs has been established. DBs can be responsible for the turbulent dynamics~\cite{Turbulence}. It was shown numerically that transition from ballistic to normal heat conduction is a consequence of presence of DBs in a nonlinear chain~\cite{Coolig12}. DBs can assist energy transfer to ac-driven nonlinear chains~\cite{Kuzkin}. DBs increase (decrease) specific heat of the nonlinear chain with soft (hard) type nonlinearity on-site potential and harmonic nearest-neighbour coupling~\cite{arXiv}. 

For a chain with on-site potential, such as considered in~\cite{arXiv}, one cannot calculate the effect of DBs on thermal expansion and on elastic constants because the on-site potential elastically pins the particles of the chain to their lattice positions and the chain cannot expand under thermal fluctuations or external force. Thermal expansion and elastic constants are very important properties in the context of solid state physics and materials science. That is why, in the present study, the Fermi-Pasta-Ulam (FPU) chain is considered and these macroscopic properties are calculated together with the specific heat. For this we use the same approach as in~\cite{arXiv}, namely, we simulate the modulational instability of the zone boundary mode ($q=\pi$) which leads first to energy localization in the form of long-lived DBs and subsequent transition to thermal equilibrium~\cite{ElLatt3,Burlakov,Mirnov,Ullmann,Kosevich,Cretegny,Mi5,Mi3,Mi4,Mi1,Mi2}. The macroscopic characteristics of the chain in the regime when energy is localized on DBs are compared with that in thermal equilibrium, thus revealing the effect of DBs on those properties. We note that properties of DBs in the FPU chain have been analyzed by Flach and Gorbach in \cite{FlachGorbach}.

Overall, this study is complementary to our previous work~\cite{arXiv} done for the chain with the on-site potential. Having on-site potential one can easily switch between hard- and soft-type nonlinearity, and comparison of these two regimes was the focus of that study. However, thermal expansion and elastic constants could not be analyzed in~\cite{arXiv}. The FPU chain considered here allows calculation of thermal expansion and elastic constants but it can support only DBs with hard-type nonlinearity having frequencies above the gapless phonon spectrum. In this respect, the models with and without on-site potential are of different physical nature, they both have numerous applications in different problems and it is important to investigate both types of chains.

Our paper is organized as follows. In Sect.~\ref{SimulationSetup} the model and simulation details are described. The simulation results on modulational instability of the zone-boundary mode and macroscopic properties of the chain are presented in Sect.~\ref{Results}. Our conclusions are presented in Sect.~\ref{Conclusion}.
\section{The model and simulation setup}
\label{SimulationSetup}

\begin{figure}[h!]
\includegraphics[width=7.5cm]{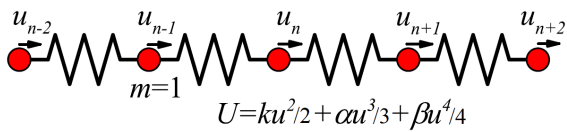}
\caption{FPU chain of harmonically coupled, unit mass point-like particles interacting with the quartic polynomial on-site potential with the nearest neighbours.}
\label{fig1}
\end{figure}

We consider the FPU chain of particles having mass $m$ (see Fig.~\ref{fig1}) described by the Hamiltonian
\begin{equation}\label{Hamiltonian}
H=K+P=\sum_n\frac{m\dot{u}_n^2}{2} 
+\sum_n U(d_{n}),
\end{equation}
where 
\begin{equation}\label{dn}
d_n=u_{n+1}-u_{n},
\end{equation}
and $K$ and $P$ are the kinetic and potential energy, $u_n^{}$ and $\dot{u}_n^{}$ are displacement from its equilibrium position and velocity of the $n$-th particle, overdot stands for differentiation with respect to time. Each particle is anharmonically coupled to their nearest neighbors:
\begin{equation}\label{PotentialHard}
U(\xi)=\frac{k}{2} \xi^2+\frac{\alpha}{3} \xi^3+\frac{\beta}{4} \xi^4,
\end{equation}
where $k$, $\alpha$ and $\beta$ are constants.

Without loss of generality, we set $m=1$ and $k=1$. We take $\beta=3$ and consider three different values for $\alpha$, namely, $\alpha\in\{0, -1/4, -1/2\}$. For $\alpha=0$ the potential is symmetric, while for the chosen negative values of $\alpha$ it is an asymmetric single-well potential.

From the Hamiltonian defined above, the following equation of motion can be derived
\begin{eqnarray}\label{EMo}
m\ddot{u}_n=
k(d_n-d_{n-1})+ \alpha(d_n^2 - d_{n-1}^2)+ \beta(d_n^3 - d_{n-1}^3).
\end{eqnarray}
The St\"ormer method of order six with the time step $\tau=10^{-3}$ is used for numerical integration of these equations. With such a time step, the relative change in total energy of the chain in a typical numerical run is not greater than $10^{-5}$.

Substituting the ansatz $u_n^{} \sim \exp [i (q n -\omega_q^{} t)]$ into Eq.~(\ref{EMo}) with $\alpha=\beta=0$, one finds the relation between wave number $q$ and frequency $\omega_q$ for the small-amplitude normal modes in the form
\begin{equation}\label{Dispersion}
\omega_q^2=\frac{2k}{m}(1-\cos q).
\end{equation}
The phonon band of the chain ranges from $\omega_{\min}^{}=0$ for $q=0$ to $\omega_{\max}^{}=2$ for $q=\pm \pi$.

Here, a chain of $N=2048$ particles is considered. Test runs with larger number of particles produced nearly the same results.

Initial conditions are set in the form of the zone-boundary mode ($q=\pi$) with the amplitude $A$,
\begin{equation}\label{ZBmode}
u_n^{}=A\sin(\pi n-\omega_{\max}^{}t).
\end{equation}
If the amplitude $A$ is not too small, then this mode is modulationally unstable. At $t=0$, all the particles have the same energy but the instability entails energy localization which can be characterized by the localization parameter 
\begin{equation}\label{Localiz}
L=\frac{\sum e_n^2}{\Big(\sum e_n^{}\Big)^2},
\end{equation}
where 
\begin{equation}\label{en}
e_n^{}=\frac{m\dot{u}_n^2}{2}+\frac{1}{2}U(d_{n})+\frac{1}{2}U(d_{n-1}),
\end{equation}
is the energy of $n$-th particle.

We define temperature as the averaged kinetic energy per atom,
\begin{equation}\label{Ken}
T=\bar{K}=\frac{1}{N}\sum_n\frac{m\dot{u}_n^2}{2}.
\end{equation}
The Boltzmann constant here is set to be equal to 2. Heat capacity of the chain is defined as 
\begin{equation}\label{HeatCap}
C=\lim_{\Delta T\to 0} \frac{\Delta H}{\Delta T},
\end{equation}
where $\Delta H$ is the increment in energy of the chain and $\Delta T$ is the corresponding temperature increment. The specific heat capacity (or simply specific heat) is the heat capacity per particle. Periodic boundary conditions are used meaning that the specific heat at constant volume is calculated.

In our simulations, total energy $H$ is conserved, so that Eq.~(\ref{HeatCap}) cannot be used. We characterize the specific heat of the chain at constant volume by the ratio
\begin{equation}\label{HeatCapHere}
c_V^{}=\frac{\bar{H}}{\bar K},
\end{equation}
where $\bar H$ ($\bar K$) is the total (kinetic) energy of the chain per atom. In linear systems $\bar H=2\bar K$ and $c_V^{}=2$. Due to nonlinearity, the kinetic energy can differ from the potential energy resulting in deviation of $c_V^{}$ from this value. Note that the relation between total to kinetic energy ratio and specific heat has been discussed in a number of works~\cite{Ebeling1,Ebeling2}, justifying the use of Eq.~(\ref{HeatCapHere}). 

Stress in the chain is calculated as follows
\begin{equation}\label{StressEq}
\sigma=\frac{1}{N}\sum_{n=1}^N(kd_n + \alpha d_n^2 + \beta d_n^3 ).
\end{equation}
Modulus of elasticity of the chain is defined by
\begin{eqnarray}\label{ModulusEq}
E=\frac{1}{N}\sum_{n=1}^N\Big[(kd_n + \alpha d_n^2 + \beta d_n^3 )(1+d_n)+  \nonumber \\
 (k + 2\alpha d_n + 3\beta d_n^2 )(1+d_n)^2
\Big].
\end{eqnarray}

In Sect.~\ref{Results}, the time evolution of localization parameter, specific heat, stress in the chain and Young's modulus of the chain have been calculated. These macroscopic characteristics have been compared in the regime of energy localization by DBs with those in thermal equilibrium.

\section{Modulational instability}
\label{Results}

We excite the zone-boundary mode Eq.~(\ref{ZBmode}) in the chain using various amplitudes $A$. From the numerical integration of Eq.~(\ref{EMo}) we find the change in the localization parameter (\ref{Localiz}), specific heat (\ref{HeatCapHere}), stress (\ref{StressEq}) and Young's modulus (\ref{ModulusEq}).

\subsection{Energy localization}
\label{Localization}

\begin{figure}[h!]
\includegraphics[width=8.5cm]{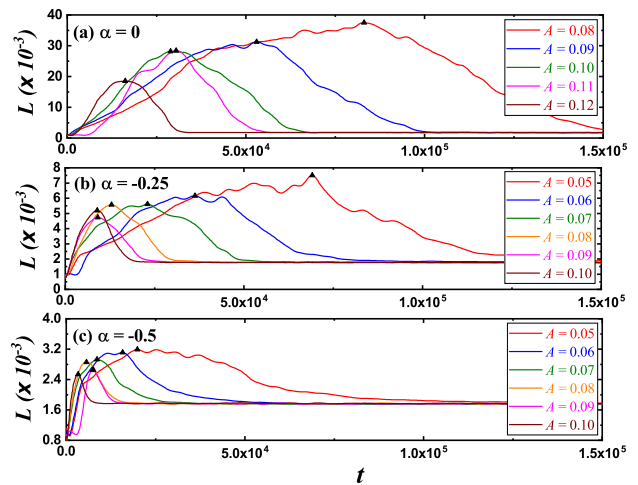}
\caption{Variation of the localization parameter as the function of time for various amplitudes of the initially excited zone-boundary mode, $A$. For all cases, at $t=0$ the localization parameter is $L=1/N=0.49 \times 10^{-3}$. There is an increase in $L$ as a consequence of energy localization on DBs due to modulational instability. Then as the DBs gradually radiate energy, $L$ decreases and eventually the system reaches thermal equilibrium with $L$ oscillating near the value of $1.8\times 10^{-3}$. The points of maxima of $L$ are marked with triangles.}
\label{fig2}
\end{figure}

Figure \ref{fig2} shows the localization parameter as a function of time for different amplitudes $A$ and three values of the potential asymmetry parameter $\alpha$. For all the curves, the localization parameter is minimal at $t=0$ that is $L=1/N=0.49\times 10^{-3}$. It increases gradually as a consequence of appearance of modulational instability that leads to formation of DBs and, hence, energy localization. Finally, the system comes to thermal equilibrium and the DBs slowly radiate their energy, the parameter $L$ gradually decreases and oscillates near $1.8\times 10^{-3}$.

\begin{figure}[h!]
\includegraphics[width=8.5cm]{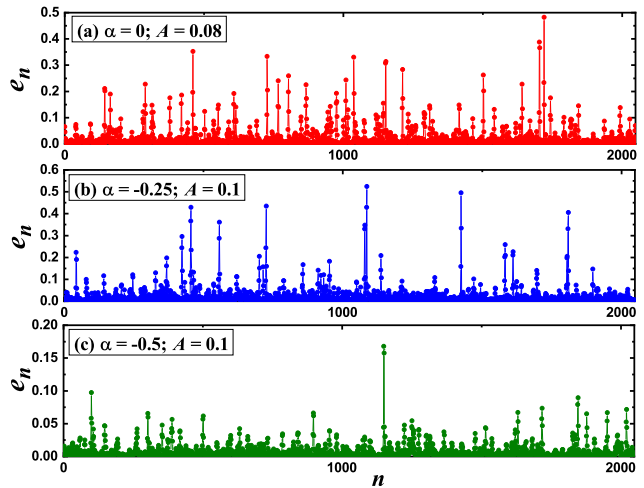}
\caption{Total energies of particles at the moment of maximal localization parameter for different values of the potential asymmetry parameter $\alpha$. In (a) the result is given for $A=0.08$ and in (b) and (c) for $A=0.1$. The results for other studied values of $A$ look qualitatively similar.}
\label{fig3}
\end{figure}

In Fig.~\ref{fig3} the energy distribution over the FPU chain is shown for various values of the potential asymmetry parameter $\alpha$ from $\alpha=0$ in (a) to $\alpha=-0.5$ in (c) at the time when localization parameter is maximal; here we see sets of highly localized DBs.

\begin{figure}[h!]
\includegraphics[width=8.5cm]{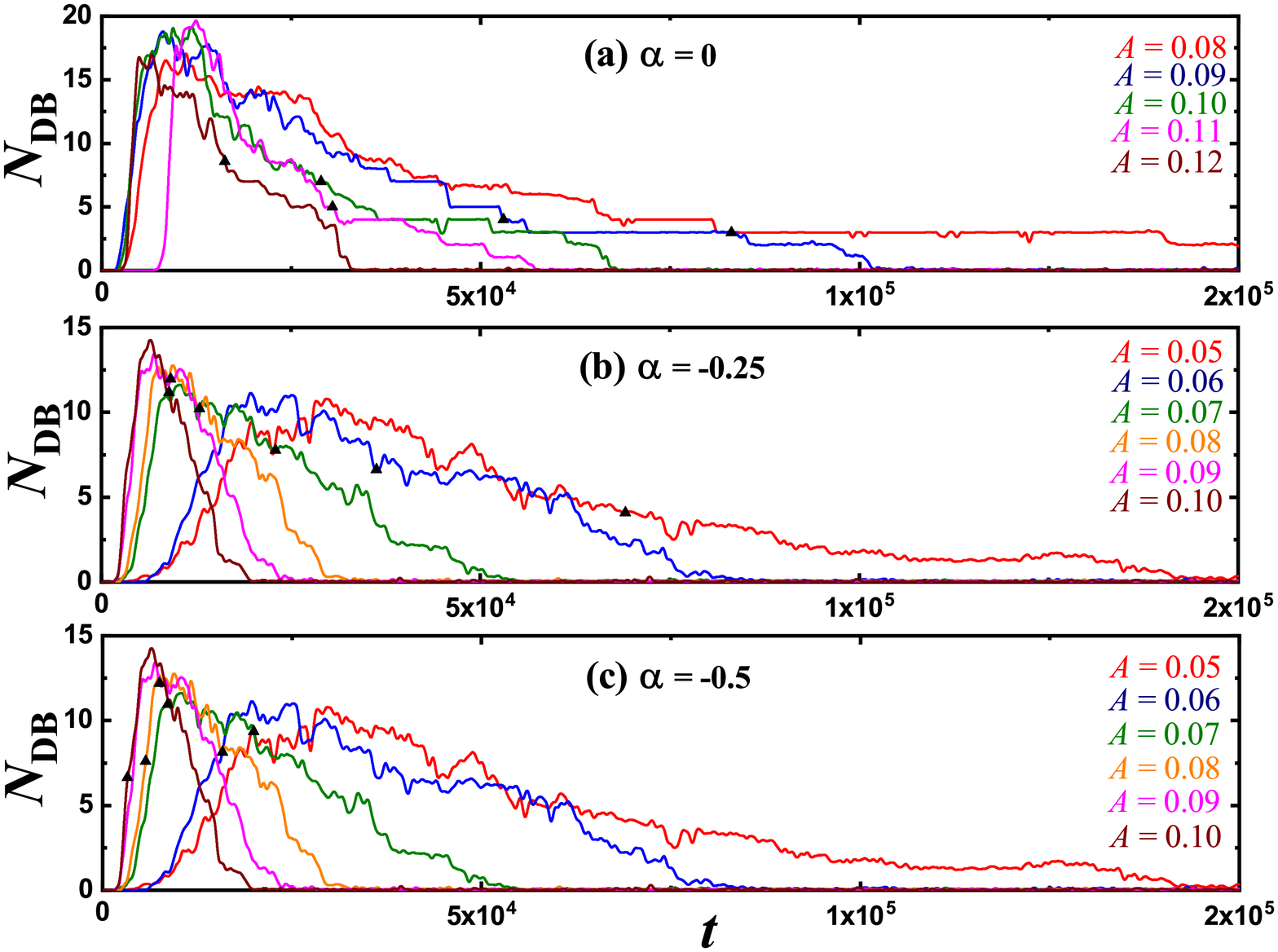}
\caption{Number of DBs as the function of time for various amplitudes of the initially excited zone-boundary mode at different values of the potential asymmetry parameter $\alpha$. Triangles indicate the points of maximal localization parameter.}
\label{fig4}
\end{figure}

\begin{figure}
\includegraphics[width=8.5cm]{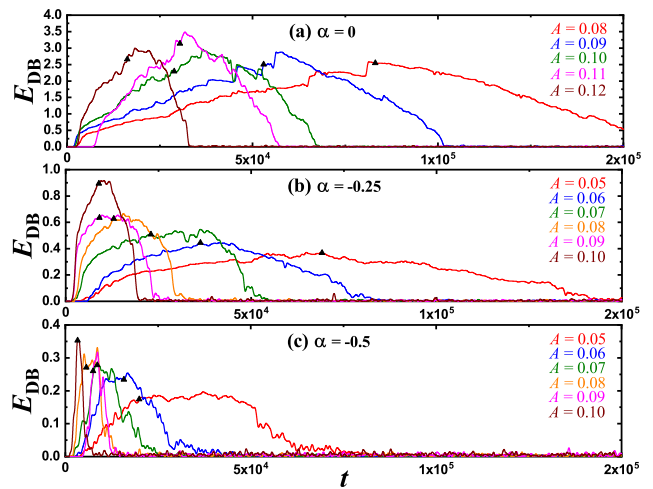}
\caption{Average DB energy as the function of time for various amplitudes of the initially excited zone-boundary mode at different values of the potential asymmetry parameter $\alpha$. Triangles indicate the points of maximal localization parameter.}
\label{fig5}
\end{figure}

The time evolution of the number of DBs produced, $N_{\rm DB}$, and the corresponding average DB energy, $E_{\rm DB}$, are shown in Fig.~\ref{fig4} and Fig.~\ref{fig5}, respectively, for different values of the zone-boundary mode amplitudes $A$ and for three values of the potential asymmetry parameter $\alpha$. In these plots, by triangles we indicate the points when the localization parameter $L$ reaches its maximum. It can be seen from Fig.~\ref{fig4} that the maximal number of DBs for $\alpha=0$ very weakly depends on $A$ and for the negative values of $\alpha$ maximal values of $N_{\rm DB}$ are somewhat greater for larger $A$. The same can be said about the maximal DB energy: it weakly depends on $A$ for $\alpha=0$ and increases with $A$ for negative $\alpha$. 

Kosevich and Kovalev have derived the criterion for the existence of DBs in FPU chain in the following form~\cite{KosevichKov}
\begin{equation}\label{InstabCriterion}
\frac{3k\beta}{4\alpha^2}>1.
\end{equation}
For our choice of model parameters, this criterion is violated for $|\alpha|>3/2$. For the values of $\alpha$ used in our simulations the criterion Eq.~(\ref{InstabCriterion}) is fulfilled but for increasing asymmetry of the potential, the conditions for the existence of DBs deteriorate. This explains the general trend seen in Fig.~\ref{fig2} to Fig.~\ref{fig5}, that the number of chaotic DBs and their energy decrease with increasing $|\alpha|$.

The zone-boundary mode Eq.~(\ref{ZBmode}) is unstable for the amplitude greater a threshold value, $A>A^*$, which depends on the potential asymmetry parameter $\alpha$. For larger asymmetry of the potential $A^*$ is smaller. That is why in Fig.~\ref{fig2} to Fig.~\ref{fig5} and in what follows we take $A\ge 0.08$ for $\alpha=0$ and $A\ge 0.05$ for $\alpha=-0.25$ and $-0.5$. For $\alpha=0$ and $A=0.05$ the zone-boundary mode is stable and it does not split into DBs.

From Fig.~\ref{fig4} one can also see that $N_{\rm DB}$ reaches its maximum well before the localization parameter $L$ becomes maximal for the case of $\alpha=0$ and $\alpha=-0.25$, but when $\alpha=-0.5$, first $L$ reaches a maximal value and then $N_{\rm DB}$. As for $E_{\rm DB}$, as shown in Fig.~\ref{fig5}, the time when it attains maximal value is close to the time when $L$ is maximal.

It is also worth pointing out that the maximal number of DBs, i.e. $N_{\rm DB}$, weakly depends on the values of $\alpha$ (see Fig.~\ref{fig4}), whereas the maximal DB energy, $E_{\rm DB}$, rapidly decreases with increasing asymmetry of the potential (see Fig.~\ref{fig5}). It can be deduced that the wavelength of modulational instability and hence, the number of energy localization centers, weakly depend on $\alpha$, but the above mentioned deterioration of DB existence condition with increasing $|\alpha|$ mostly affects the average DB energy. 

\subsection{Total to kinetic energy ratio (specific heat)}
\label{SpecificHeat}

\begin{figure}[h!]
\includegraphics[width=8.5cm]{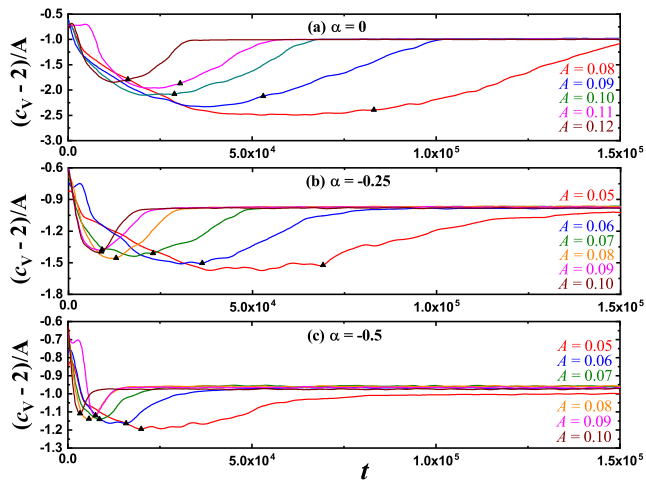}
\caption{Specific heat normalized with respect to the zone-boundary mode amplitude $A$ as a function of time for various values of $A$ at three different values of the potential asymmetry parameter $\alpha$. The black triangular markers indicate the corresponding values at maximal localization parameter $L$. It can be seen that specific heat is close to minimum when DBs are in the system and it increases as the system approaches thermal equilibrium.}
\label{fig6}
\end{figure}

The variation of specific heat with respect to time is plotted for the various mode amplitudes $A$ in Fig.~\ref{fig6} for three different values of the potential asymmetry parameter $\alpha$. We actually present the deviation of specific heat from its theoretical value of 2 for the linear system, normalized by $A$. The black triangular markers represent the values at which the localization parameter is maximal. From the comparison of Fig.~\ref{fig2} and Fig.~\ref{fig6}, we can observe that the specific heat is minimal when the the localization parameter is close to its maximum. The effect is most prominent for symmetric potential $(\alpha=0)$ and it becomes weaker with increasing potential asymmetry. Thus, the minimal value of $(c_V^{}-2)/A$ for $\alpha=0$ is about $-2.5$, while for $\alpha=-0.25$ and $-0.5$ it is $-1.5$ and $-1.2$, respectively. This correlates with the fact that maximal localization parameter reduces with increasing $|\alpha|$ (see Fig.~\ref{fig2}). The specific heat increases during the transition to thermal equilibrium. From this, we conclude that the specific heat of the chain is reduced by the DBs resulting from the modulational instability of the zone-boundary mode. This can be explained quite simply as in our system with hard type anharmonicity, the DB frequency increases with its amplitude. Increase in the oscillation frequency results in an increase of particle velocities and thus, in their kinetic energies. Kinetic energy is in the denominator of Eq.~(\ref{HeatCapHere}) (which means there is an inverse proportionality between $c_V^{}$ and $K^{}$) and hence its increase results in a decrease of $c_V^{}$. Analogously, for the case of soft type anharmonicity, DB frequency decreases with its amplitude and the effect is just opposite, i.e., DBs increase the specific heat~\cite{arXiv}.

\subsection{Stress}
\label{Stress}

\begin{figure}[tbh!]
\includegraphics[width=8.5cm]{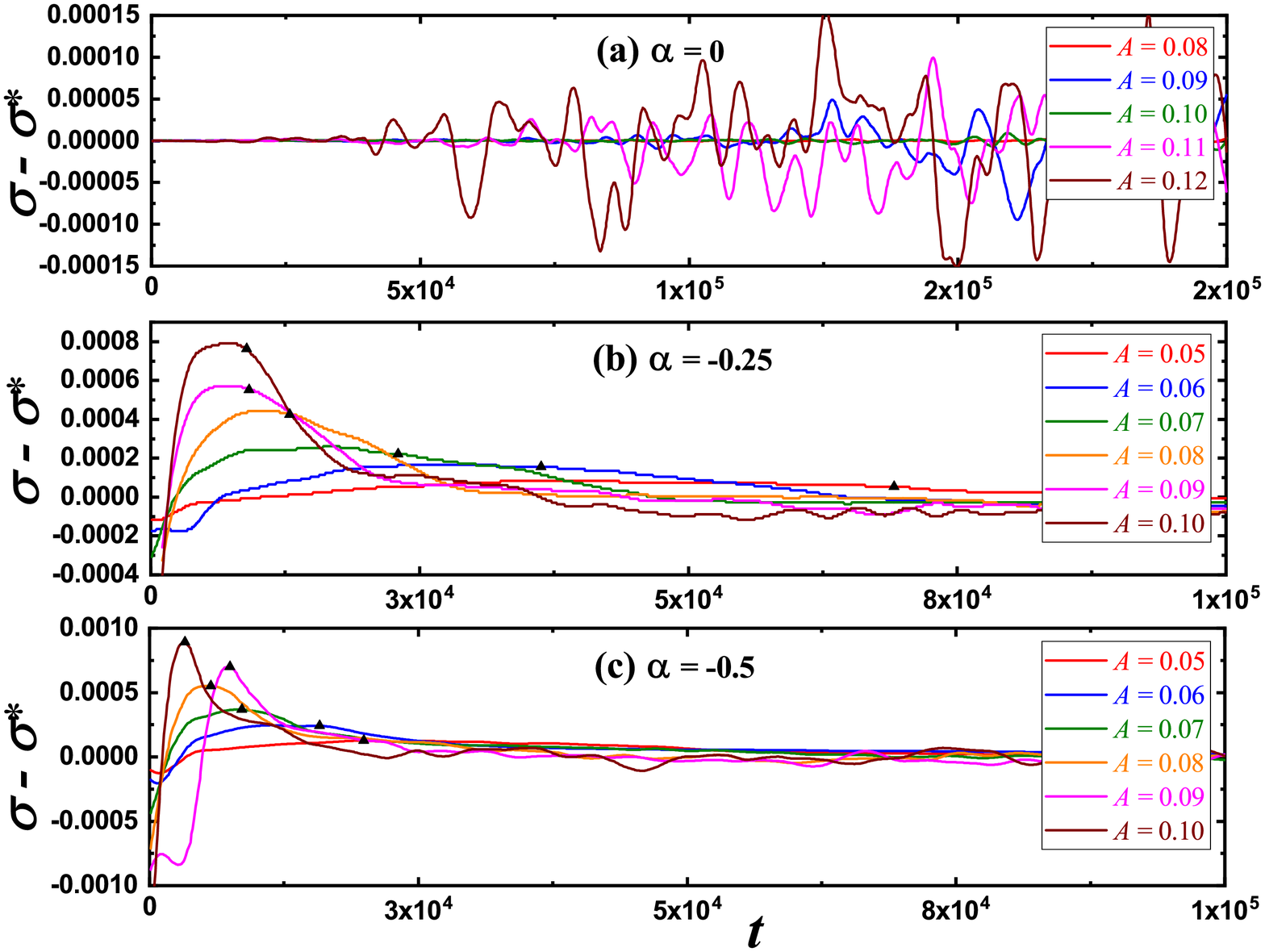}
\caption{Stress as the function of time for various amplitudes of the initially excited zone-boundary mode at different values of the potential asymmetry parameter ($\alpha$). The triangular markers indicate the corresponding values at maximal localization. The $\sigma$ values have been normalized by subtracting the corresponding thermal equilibrium values $\sigma^*$ to get the results in a comparable range.}
\label{fig7}
\end{figure}

The time-dependence of stress in the linear chain is plotted for the various mode amplitudes $A$ in Fig.~\ref{fig7} for different values of the potential asymmetry parameter: (a) $\alpha=0$, (b) $\alpha=-1/4$, and (c) $\alpha=-1/2$. We present here the deviation of stress $\sigma$ from its value in thermal equilibrium,
\begin{equation}\label{sigmas}
\sigma^*=\frac{1}{t_2-t_1}\int_{t_1}^{t_2}\sigma dt,
\end{equation}
where $t_1$ is the time when system reaches the state of thermal equilibrium with stress oscillating near a constant value, and $t_2-t_1=10^4$ is the sufficiently long time of averaging. 

\begin{figure}[tbh!]
\includegraphics[width=8.5cm]{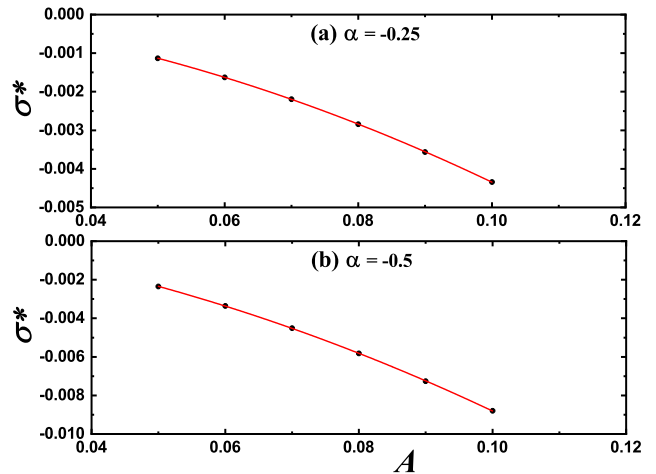}
\caption{The stress at thermal equilibrium as the function of amplitude of the initially excited zone-boundary mode for different values of the potential asymmetry parameter ($\alpha$).}
\label{fig8}
\end{figure}

For the symmetric potential ($\alpha=0$) there is no thermal expansion and $\sigma^*=0$. For negative values of $\alpha$ the dependence of $\sigma^*$ on $A$ is given in Fig.~\ref{fig8}. In the limit of very small $A$, i.e., for the linear regime with small amplitude phonons, the stress is zero. For greater asymmetry, the absolute value of $\sigma^*$ is larger for the same value of $A$ [cf. (a) and (b)]. For $\alpha=-0.5$ the values of $\sigma^*$ are nearly doubled as compared to those for $\alpha=-0.25$, revealing nearly linear dependence of $\sigma^*$ on $\alpha$.

Coming back to Fig.~\ref{fig7}, once again according to the position of the triangular markers or from the comparison of Fig.~\ref{fig2} and Fig.~\ref{fig7}, it can be seen in (b) and (c) that the stress is maximal when the localization parameter is close to its maximum. During the transition to thermal equilibrium, the stress in the chain decreases. Maximal value of $\sigma-\sigma^*$, observed for $\alpha=-0.25$, is 0.0008 and it is one order of magnitude greater for $\alpha=-0.5$. This is understandable taking into account the this effect is due to the asymmetry of the potential and, naturally, it becomes stronger for increasing asymmetry.

Note that negative (compressive) stress appears in the system because its thermal expansion is suppressed by the use of periodic boundary conditions. In the chain with free ends, for negative $\alpha$, thermal expansion at zero stress will be observed. 

From this, we conclude that the DBs increase the stress in the FPU chain with periodic boundary conditions and in the case of free boundary conditions they will produce thermal expansion at zero stress.

\begin{figure}[th!]
\includegraphics[width=8.5cm]{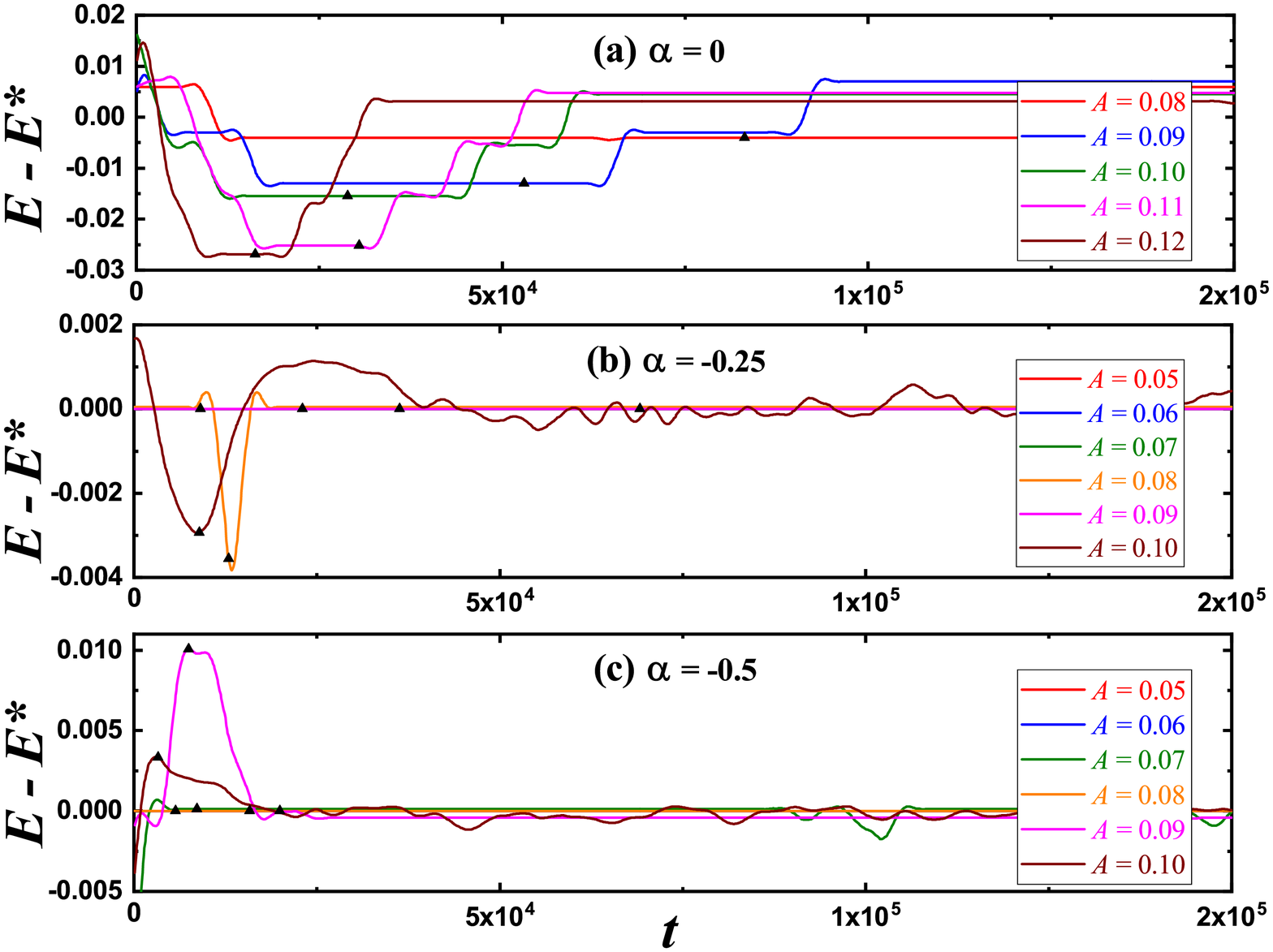}
\caption{Modulus of elasticity as the function of time for various amplitudes of the initially excited zone-boundary mode at different values of the potential asymmetry parameter ($\alpha$). The triangular markers indicate the corresponding values at maximal localization. The values of $E$ have been normalized by subtracting the corresponding thermal equilibrium values $E^*$ to get the results in a comparable range.}
\label{fig9}
\end{figure}

\subsection{Modulus of elasticity}
\label{Modulus}

\begin{figure}[h!]
\includegraphics[width=8.5cm]{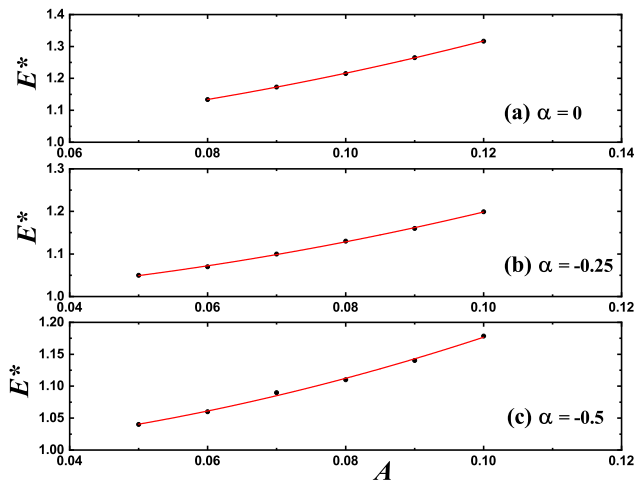}
\caption{The Young's Modulus of elasticity at thermal equilibrium as the function of amplitudes of the initially excited zone-boundary mode for different values of the potential asymmetry parameter ($\alpha$).}
\label{fig10}
\end{figure}

The time-dependence of the Young's modulus of elasticity $E$ for the FPU chain is plotted for the various mode amplitudes $A$ in Fig.~\ref{fig9} for (a) $\alpha=0$, (b) $\alpha=-1/4$, and (c) $\alpha=-1/2$. Again, the difference between $E$ and its value in thermal equilibrium, $E^*$, is given. The latter is calculated as
\begin{equation}\label{Es}
E^*=\frac{1}{t_2-t_1}\int_{t_1}^{t_2}E dt,
\end{equation}
where $t_1$ is the time when system reaches the state of thermal equilibrium with modulus oscillating near a constant value, and $t_2-t_1=10^4$ is the time of averaging. The values of $E^*$ for different initial mode amplitudes are plotted in Fig.~\ref{fig10} for (a) $\alpha=0$, (b) $\alpha=-1/4$, and (c) $\alpha=-1/2$.

From the comparison of Fig.~\ref{fig2} and Fig.~\ref{fig9}, it can be seen that the modulus of elasticity is minimal when the localization parameter is maximal for $\alpha=0$ and $\alpha=-0.25$. In the limit of very small $A$ one has $E^*=1$ and the compressive rigidity of the chain increases with increasing $A$ almost equally for different values of $\alpha$. 

From Fig.~\ref{fig9}, it can be seen that the effect of DBs on the Young's modulus is more pronounced for the symmetric potential ($\alpha=0$). In this case, $E-E^*$ is minimal when DBs are in the system and it increases while the system approaches thermal equilibrium. From this, we conclude that the resulting DBs decrease the modulus of elasticity of the FPU chain with symmetric potential. 

The effect of the Young's modulus reduction by DBs is much weaker for $\alpha=-1/4$ and this trend gets even reversed for $\alpha=-1/2$, i.e., in this case the modulus of elasticity is maximal when the localization parameter is maximal and decreases in thermal equilibrium. Thus, no definite conclusion can be made about the effect of DBs on the Young's modulus of the FPU chain with the asymmetric potential since it can be qualitatively different for different values of the asymmetry parameter.

This non-monotonous dependence of $E-E^*$ on $\alpha$ can be explained by the fact that $E$ depends not only on $dU/d\xi$, but also on $d^2U/d\xi^2$, see Eq.~(\ref{ModulusEq}), and thus, it is defined by the interplay between these two different quantities. In contrast, stress in the chain depends only on $dU/d\xi$, see Eq.~(\ref{StressEq}), and it shows a monotonous dependence on $\alpha$.

\section{Conclusions}
\label{Conclusion}

In this study, the effect of DBs on different macroscopic properties of $\alpha$-$\beta$-FPU chain was discussed. The properties such as total to kinetic energy ratio (related to specific heat), internal stress (related to thermal expansion) and Young's modulus were measured during the transition from modulationally unstable zone-boundary mode through the regime with high energy localization on DBs to thermal equilibrium. 

It was found that for the chain with any set of parameters, specific heat is reduced by DBs. This is due to the hard-type anharmonicity of the chain with DBs having greater vibrational frequencies for greater amplitudes. In the regime of energy localization by DBs, vibration frequencies increase, and this leads to an increase in particles velocities and thus, their kinetic energies. Then, according to Eq.~(\ref{HeatCapHere}), an increase of kinetic energy in the DB regime results in the reduction of the specific heat. In the chains with soft-type anharmonicity DB frequency drops with its amplitude and the effect is opposite, i.e., DBs increase the specific heat~\cite{arXiv}. 

Internal stress in the chain with symmetric potential ($\alpha=0$) does not appear since thermal expansion is observed only for asymmetric potentials. For negative $\alpha$ values considered in this study, negative (compressive) stress appears in the chain with periodic boundary conditions that suppress free thermal expansion. If free boundary conditions were used, thermal expansion of the chain at zero stress would be observed. The compressive stress is greater in the regime with DBs and thus, DBs increase thermal expansion of the FPU chain with negative $\alpha$.

In the chain with symmetric potential, DBs reduce the Young's modulus, but with increasing asymmetry the effect gets weaker and for $\alpha=-1/2$ it becomes even reverse, i.e, an increase in modulus of elasticity is observed. 

The results obtained in this work for FPU chain, which does not feature an on-site potential, complement the previous study~\cite{arXiv} for the chain with the on-site potential. In~\cite{arXiv} the analysis of thermal expansion and elastic constants was in principle impossible and only heat capacity was analyzed. Here we were able to calculate all these important properties. On the other hand, the chain considered in~\cite{arXiv}, depending on the parameters, supports DBs with hard- and soft-type nonlinearity, while FPU chain supports only hard-type nonlinearity DBs.

Our results help to interpret the results of experimental studies on the effect of DBs on macroscopic properties of crystals~\cite{UthermalExp,UheatCapac}. In particular, in the work~\cite{UheatCapac} it has been suggested that heat capacity of $\alpha$-uranium increases at high temperatures due to excitation of DBs. However, phonon spectrum of $\alpha$-uranium is gapless, it supports hard-type anharmonicity DBs~\cite{M4}, which can only decrease heat capacity, as it follows from the results presented here and in the work~\cite{arXiv}.

The effect of DBs on macroscopic properties of nonlinear lattices of higher dimension and of real crystals could become a subject of future study.

\begin{acknowledgments}
E.A.K.\ acknowledges the support of the Grant of the President of the Russian Federation for State support of young Russian scientists (No.\ MD-3639.2019.2). V.A.G.\ acknowledges the support of the MEPhI Academic Excellence Project (Contract No.~02.a03.21.0005, 27.08.2013). The work of S.V.D.\ was supported by the Russian Foundation for Basic Research, Grant No.\ 19-02-00971. The work of A.M.K.\ and V.A.K.\ was financially supported by the Russian Science Foundation, Grant No.~18-11-00201.

V.A.G.\ also thanks Tatiana Gani for her help in editing the manuscript.
\end{acknowledgments}



\end{document}